\documentclass[a4paper,11pt]{article}
\usepackage{lineno}
\usepackage{amssymb}
\usepackage{amsfonts,amsbsy,graphics,epsfig}
\usepackage{graphicx}
\usepackage{amsmath}
\usepackage{color}
\usepackage{multirow}
\usepackage{arydshln}

\usepackage[dvipsnames]{xcolor}
\usepackage{wasysym}
\usepackage{natbib}
\usepackage{cleveref}
\usepackage{adjustbox}
\modulolinenumbers[1]

\usepackage{stmaryrd} 
\usepackage{dsfont} 

\begin{document}
\begin{center}
%
%

\textbf{\Large{Detecting spatial clusters in functional data: new scan statistic approaches}}\medskip \\
Camille Frévent$^1$, Mohamed-Salem Ahmed$^{1}$ , Matthieu Marbac$^2$ and Michaël Genin$^1$  \\
$^1$Univ. Lille, CHU Lille, ULR 2694 - METRICS: Évaluation des technologies de santé et des pratiques médicales, F-59000 Lille, France.\\ 
$^2$Univ. Rennes, Ensai, CNRS, CREST - UMR 9194, F-35000 Rennes, France.
\end{center}	
\begin{center}
	\rule{1\linewidth}{.9pt}
\end{center}
\noindent	
\textbf{\Large Abstract}\medskip \\
\noindent We have developed two scan statistics for detecting clusters of functional data indexed in space. The first method is based on an adaptation of a functional analysis of variance and the second one is based on a distribution-free spatial scan statistic for univariate data. In a simulation study, the distribution-free method always performed better than a nonparametric functional scan statistic, and the adaptation of the anova also performed better for data with a normal or a quasi-normal distribution. Our methods can detect smaller spatial clusters than the nonparametric method. Lastly, we used our scan statistics for functional data to search for spatial clusters of abnormal unemployment rates in France over the period 1998-2013 (divided into quarters).\\

\begin{center}
\rule{1\linewidth}{.9pt}
\end{center}
\noindent \section{Introduction}
Spatial cluster detection has been studied for many years. The goal is usually to develop new tools capable of detecting the aggregation of spatial sites that behave ``differently'' from other sites. There are three main categories of cluster detection tests: (i) global clustering tests, designed to determine whether some spatial observations tend to aggregate (but without locating the corresponding cluster), (ii) focused cluster detection tests, which screen for the presence of one or more clusters around a fixed focus (but which require prior knowledge, since a focus determined on the data generates pre-selection bias), and (iii) non-focused cluster detection tests, designed to detect statistically significant spatial clusters in the absence of any a priori information about the latter’s location. Spatial scan statistics are well-known members of the third category and have been applied in many fields of applied research, such as epidemiology \citep{epidemio1, epidemio2, genin2020fine}, environmental sciences \citep{social2,social1}, and geology \citep{geology}. \\

\noindent Scan statistics were first introduced by Naus for the one-dimensional case \cite{naus65one} and then the two-dimensional case \cite{naus65two}. In fact, spatial scan statistics can be viewed as the extension of two-dimensional scan statistics to spatial data. This approach was initially suggested by \cite{spatialdisease} and \cite{spatialscanstat} for Bernouilli and Poisson models, respectively. The researchers used a method based on the likelihood ratio and Monte-Carlo testing to detect significant clusters of various sizes and shapes. Following on from Kulldorff’s initial work, several other researchers have adapted spatial scan statistics to other spatial data distributions, such as ordinal \citep{jung2007spatial}, normal \citep{normalkulldorff}, exponential \citep{huang2007spatial}, and Weibull models \citep{bhatt2014spatial}.
\cite{kulldorffmulti} developed a parametric scan statistic for application to multivariate spatial data. However, this method is based on the combination of independent univariate scan statistics and fails to take account of the correlations between variables. This obstacle was circumvented by  \cite{a_multivariate_gaussian}, who developed a spatial scan statistic based on a likelihood ratio and a multivariate normal probability model that takes account of the correlations between variables.\\

\noindent Thanks to progress in sensing and data storage capacity, data are increasingly being measured continuously over time. This led to the introduction of functional data analysis (FDA) by \cite{ramsaylivre}. A considerable amount of work has gone into the representation, exploration and modelling of functional data and the adaptation of classical statistical methods to the functional framework \citep[e.g. principal component analysis,][]{fpc_boente,  fpc_berrendero} and regression \citep{reg_cuevas, reg_ferraty, reg_chiou}. Nonparametric methods have also been developed for FDA \citep[for a review, see][]{ferratybook}.  FDA is thus an active research topic with potential applications in many fields. \\
In the field of spatial scan statistics, the application of a ``naive'' univariate approach (based on the time-averaged data in each spatial location over the studied period) would lead to a huge loss of information. A multivariate method that considers each observation time point as a variable would be face with high dimensionality and high correlation issues. Recently, \cite{wilco_cucala} developed a nonparametric approach based on a functional Wilcoxon-Mann-Whitney test. To the best of our knowledge, a parametric scan statistic for functional data has not previously been suggested. We reasoned that since an analysis of variance (ANOVA) for functional data has been described by \cite{anova_functional} it should be possible to develop a scan procedure based on this test. Then in the last few years, statistical tests for high-dimensional data were developed by resuming the information after the computation of the statistics for each component of the data \citep{hd_manova}. Hence we proposed a scan statistic based on the combination of this approach and the distribution-free scan statistic approach proposed by \cite{a_distribution_free}. \\

\noindent Here, we describe a parametric spatial scan statistic for functional data based on a functional ANOVA and another one based on the distribution-free scan statistic. The methodology is presented in Section \ref{sec:method}. In Section \ref{sec:simulation}, the new approaches’ performances in a simulation study are described and compared with that of the \cite{wilco_cucala} method. Our methods’ application to a real dataset is presented in Section \ref{sec:realdata}. Lastly, the results of this work are discussed in Section \ref{sec:discussion}. \\

\section{Methodology}
\label{sec:method}
\subsection{General principle}
\noindent Let $s_1, \dots, s_n$ be $n$ non-overlapping locations of an observation domain $S \subset \mathbb{R}^2$ and $X_1, \dots, X_n$ be the observations of $X$ in $s_1, \dots, s_n$. Hereafter, all observations are considered to be independent; this is a classical assumption in scan statistics. Spatial scan statistics are designed to detect spatial clusters and to test their statistical significance. Hence, one tests a null hypothesis $\mathcal{H}_0$ (the absence of a cluster) against a composite alternative hypothesis $\mathcal{H}_1$ (the presence of at least one cluster $w \subset S$ presenting abnormal values of $X$).\\

\noindent When $X \in \mathbb{R}$, parametric univariate scan statistics can detect aggregates of locations in which the mean of $X$ is higher or lower than the mean of $X$ outside. To this end, they often use likelihood ratios \citep{normalkulldorff}. In such a case, a cluster $w$ can be defined by $\mathbb{E}[X_i\mid s_i\in w] = \mathbb{E}[X_i\mid s_i\notin w] + \Delta \text{ where } \Delta \neq 0$.\\

\noindent In the same way, in the multivariate case ($X \in \mathbb{R}^d, \ d \ge 2$), parametric spatial scan statistics also use likelihood ratios \citep{kulldorffmulti, a_multivariate_gaussian} to detect aggregates of locations in which the values of the random vector $X$ are abnormally high or low. Hence, a cluster $w$ in this framework can be characterized by $\mathbb{E}[X_i\mid s_i\in w] = \mathbb{E}[X_i\mid s_i\notin w] + \Delta \text{ where } \Delta = (\Delta_1, \dots, \Delta_d)^\top \neq 0$, and for all $i \in \llbracket 1 ; d \rrbracket, \ \Delta_i \ge 0$ or for all $i \in \llbracket 1 ; d \rrbracket, \ \Delta_i \le 0$. \\

\noindent In the functional framework, $\{ X(t), \ t \in \mathcal{T} \}$ is a real-valued stochastic process where $\mathcal{T}$ is an interval of $\mathbb{R}$. For functional data, and in line with Dai \& Genton’s work \citep{functional_outlying} on magnitude outlyingness and shape outlyingness, two types of clusters can be distinguished: \textit{magnitude clusters} and \textit{shape clusters}. Thus, by analogy with the univariate and multivariate definitions, a \textit{magnitude cluster} $w$ can be defined as: 
\begin{equation}
\forall t \in \mathcal{T}, \ \mathbb{E}[X_i(t)\mid s_i\in w] = \mathbb{E}[X_i(t)\mid s_i\notin w] + \Delta(t),
\end{equation}
where $\Delta$ is of constant sign, and non-zero over at least one sub-interval of $\mathcal{T}$. 

\noindent In the same way, a \textit{shape cluster} $w$ can be defined as: 

\begin{equation}
\forall t \in \mathcal{T}, \ \mathbb{E}[X_i(t)\mid s_i\in w] = \mathbb{E}[X_i(t)\mid s_i\notin w] + \Delta(t)
\end{equation}
where $\Delta$ is not constant almost everywhere.

\noindent \cite{cressie} defined a spatial scan statistic as the maximum of a concentration index over a set of potential clusters $\mathcal{W}$. The spatial scan statistic thus depends on the choice of the concentration index and of $\mathcal{W}$. In the following and without loss of generality, $\mathcal{W}$ is composed of circular clusters (as introduced by \cite{spatialscanstat}), with $|w|$ denoting the number of sites in $w$:  
\begin{equation}
\mathcal{W} = \{ w_{i,j} \ / \ 1 \le |w_{i,j}| \le \frac{n}{2}, \ 1 \le i,j \le n \},
\end{equation}
where $w_{i,j}$ is the disc centred on $s_i$ that passes through $s_j$: a cluster cannot cover more than 50\% of the studied region. It should be noted that researchers have suggested many other shapes, such as elliptical clusters \citep{elliptic}, rectangular clusters \citep{rectangular} or graph-based clusters \citep{cucala_graph}. \\

\noindent From now on, we focused to the case of functional univariate data. In subsection \ref{subsec:param} we proposed a parametric scan statistic and a distribution-free one in subsection \ref{subsec:dfree}.

\subsection{A parametric spatial scan statistic for univariate functional data} \label{subsec:param}

\noindent As in the univariate and multivariate frameworks, the parametric concentration index defined here will compare the mean functions between the potential clusters and the outside.

\noindent In this subsection, the process $X$ is supposed to take values in the space $L^2(\mathcal{T}, \mathbb{R})$ of real valued, square-integrable functions on $\mathcal{T}$. 

\noindent \cite{anova_functional} and \cite{comparisonanova} adapted the classical ANOVA F-statistic for $L^2$ processes. Without loss of generality, and considering two independent samples of trajectories drawn from two $L^2$ processes $X_{g_1}$ and $X_{g_2}$ in two groups $g_1$ and $g_2$, the test compares the mean functions $\mu_{g_1}$ and $\mu_{g_2}$ where $\mu_{g_i}(t) = \mathbb{E}[X_{g_i}(t)]$, $i=1,2$.

\noindent Thus, in the context of cluster detection, the null hypothesis $\mathcal{H}_0$ (the absence of cluster) can be defined as follows: $\mathcal{H}_0: \forall w \in \mathcal{W}, \ \mu_{w} = \mu_{w^\mathsf{c}} = \mu_S$, where $\mu_w$, $\mu_{w^\mathsf{c}}$ and $\mu_S$ stand for the mean functions in $w$, outside $w$ and over $S$, respectively. The alternative hypothesis $\mathcal{H}_1^{(w)}$ associated with a potential cluster $w$ can be defined as follows: $\mathcal{H}_1^{(w)}: \mu_{w} \neq \mu_{w^\mathsf{c}}$. Next, the functional ANOVA can be used to compare the mean function in $w$ with the mean function in $w^\mathsf{c}$ by using the following statistic: 
\begin{equation}
F_n^{(w)} = \dfrac{
|w| \ ||\bar{X}_{w} - \bar{X}||_2^2 + |w^\mathsf{c}|  \  ||\bar{X}_{w^\mathsf{c}} - \bar{X}||_2^2}{ \dfrac{1}{n-2}
\left[\sum_{j, s_j \in w} ||X_j - \bar{X}_{w}||_2^2 + \sum_{j, s_j \in w^\mathsf{c}} ||X_j - \bar{X}_{w^\mathsf{c}}||_2^2 \right]},
\end{equation}
where $\bar{X}_{g}(t) = \dfrac{1}{|g|} \sum_{i, s_i \in g} X_i(t)$ are empirical estimators of $\mu_g$ ($g \in \{w, w^\mathsf{c}\}$), $\bar{X}(t) = \dfrac{1}{n} \sum_{i = 1}^n X_i(t)$ is the empirical estimator of $\mu_S$ and $||x||_2^2 = \int_{\mathcal{T}} x^2(t) \ \text{d}t$.

\noindent Thus, $F_n^{(w)}$ is considered to be a concentration index and is maximized over the set of potential clusters $\mathcal{W}$, yielding to the following definition of the parametric functional spatial scan statistic (PFSS):
\begin{equation}
\Lambda_{\text{PFSS}} = \underset{w \in \mathcal{W}}{\max} \  F_n^{(w)}.
\end{equation}
The potential cluster for which this maximum is obtained (namely, the most likely cluster (MLC)) is therefore
\begin{equation}
\text{MLC} = \underset{w \in \mathcal{W}}{\arg \max} \  F_n^{(w)}.
\end{equation}

\subsection{A distribution-free spatial scan statistic for univariate functional data} \label{subsec:dfree}

Here, we propose to combine the distribution-free spatial scan statistic for univariate data proposed by \cite{a_distribution_free} and the max statistic of \cite{hd_manova}. Briefly, the latter proposed a new approach to the problem of the functional ANOVA by maximizing a statistic over the time. \\ 

\noindent In line with the work of \cite{a_distribution_free}, we suppose that for each time $t$, $\mathbb{V}[X_i(t)] = \sigma^2(t)$ for all $i \in \llbracket 1 ; n \rrbracket$. Then for each $t$, the concentration index proposed by \cite{a_distribution_free} to test $\mathcal{H}_0: \forall w \in \mathcal{W}, \mu_w(t) = \mu_{w^\mathsf{c}}(t) = \mu_S(t)$ is 

$$I^{(w)}(t) = \dfrac{|\bar{X}_w(t) - \bar{X}_{w^\mathsf{c}}(t)|}{\sqrt{\hat{\mathbb{V}}[\bar{X}_w(t) - \bar{X}_{w^\mathsf{c}}(t)]}}, $$
where $\bar{X}_{g}(t) = \dfrac{1}{|g|} \sum_{i, s_i \in g} X_i(t)$ are empirical estimators of $\mu_g$ ($g \in \{w, w^\mathsf{c}\}$) and \\ $\hat{\mathbb{V}}[\bar{X}_w(t) - \bar{X}_{w^\mathsf{c}}(t)] = \widehat{\sigma^2}(t) \left[ \dfrac{1}{|w|} + \dfrac{1}{|w^\mathsf{c}|} \right] $. \\
Here $\widehat{\sigma^2}(t)$ needs to be computed since it depends on $t$: $$\widehat{\sigma^2}(t) = \dfrac{1}{n-2} \left[ \sum_{i, s_i \in w} (X_i(t) - \bar{X}_w(t))^2 + \sum_{i, s_i \in w^\mathsf{c}} (X_i(t) - \bar{X}_{w^\mathsf{c}}(t))^2 \right]$$ is the pooled sample variance at time $t$. \\

\noindent So now the idea is to globalize the information by maximizing the previous quantity over the time for each potential cluster $w$, as suggested by \cite{hd_manova}:

$$ I^{(w)} = \underset{t \in \mathcal{T}}{\sup} \ I^{(w)}(t).$$

\noindent For cluster detection, as for the PFSS, the null hypothesis $\mathcal{H}_0$ (the absence of cluster) is defined as follows: $\mathcal{H}_0: \forall w \in \mathcal{W}, \ \mu_{w} = \mu_{w^\mathsf{c}} = \mu_S$. And the alternative hypothesis $\mathcal{H}_1^{(w)}$ associated with a potential cluster $w$ can be defined as follows: $\mathcal{H}_1^{(w)}: \mu_{w} \neq \mu_{w^\mathsf{c}}$. \\

\noindent $I^{(w)}$ can be considered as a concentration index and maximized over the set of potential clusters $\mathcal{W}$ yielding to the following distribution-free functional spatial scan statistic (DFFSS):

\begin{equation}
\Lambda_{\text{DFFSS}} = \underset{w \in \mathcal{W}}{\max} \  I^{(w)}.
\end{equation}
The potential cluster for which this maximum is obtained (namely, the most likely cluster (MLC)) is therefore
\begin{equation}
\text{MLC} = \underset{w \in \mathcal{W}}{\arg \max} \  I^{(w)}.
\end{equation}

\subsection{Computing the significance of the MLC} \label{subsec:significance}

\noindent Once the MLC has been detected, its statistical significance must be evaluated. However, due to the overlapping nature of $\mathcal{W}$,  the distribution of the scan statistic $\Lambda$ ($\Lambda_{\text{PFSS}}$ or $\Lambda_{\text{DFFSS}}$) does not have closed form under $\mathcal{H}_0$. 
Hence, we created a large set of random datasets by randomly permuting the observations $X_i$ in the spatial locations. This technique is called ``random labelling'' and was already used in spatial scan statistics \citep{normalkulldorff, a_multivariate_gaussian, nonparam_multi}.

\noindent Let $M$ denote the number of random permutations of the original dataset and $\Lambda^{(1)},\dots,\Lambda^{(M)}$ be the scan statistics observed in the simulated datasets. According to \cite{dwass} the p-value for $\Lambda$ with real data is estimated by
\begin{equation}
\hat{p} = \frac{1 + \sum_{m=1}^M \mathds{1}_{\Lambda^{(m)} \ge \Lambda}}{M+1}.
\end{equation}
Lastly, the MLC is considered to be statistically significant if the associated $\hat{p}$ is less than the type I error.

\section{The simulation study} \label{sec:simulation}

\noindent In a simulation study, we compared (i) the parametric functional spatial scan statistic (PFSS) $\Lambda_{\text{PFSS}}$, (ii) the distribution-free functional spatial scan statistic (DFFSS) $\Lambda_{\text{DFFSS}}$ and (iii) the nonparametric functional spatial scan statistic (NPFSS) $\Lambda_{\text{NPFSS}}$ developed by  \cite{wilco_cucala}.

\noindent As stated by \cite{wilco_cucala}, calculation of the NPFSS is very time-consuming because $|w| (n - |w|)$ terms have to be summed for each potential cluster. Although \cite{wilco_cucala} suggested a computational improvement, the number of calculations was still high. Here, we found a way of improving the NPFSS computation that significantly reduced the computing time. Details of this optimization and examples of computation time are provided in the Appendix (Section \ref{Reduction_computing_time}).

\subsection{Design of the simulation study}

\noindent Artificial datasets were generated by using the geographic locations of the 94 French \textit{départements} (county-type administrative areas; Figure \ref{fig:sitesgrid}). The location of each \textit{département} was defined by its administrative capital. A true cluster $w$ (composed of eight \textit{départements} in the Paris region; the red area, see Figure \ref{fig:sitesgrid} in Supplementary materials)  was defined and simulated for each artificial dataset.

\subsubsection{Generation of the artificial datasets}

\noindent The $X_i$ are generated using the following model  \citep[see][for more details]{two_s} and measured at 101 equally spaced times on $[0;1]$:
\begin{equation*}
\text{for each }i \in \llbracket 1 ; 94 \rrbracket,\  X_i(t) = \sin{[2\pi t^2]}^5 + \Delta(t) \mathds{1}_{s_i \in w} + \varepsilon_{i}(t), \ t \in [0;1] 
\end{equation*}
where $\varepsilon_{i}(t) = \sum_{k=1}^7 \sqrt{1.5 \times 0.2^k} (v_{i,1,k} - v_{i,2,k}) \Psi_k(t)$
and
$$\Psi_k(t) = \left\{ \begin{array}{ll}
1 & \text{ if } k = 1 \\
\sqrt{2} \sin{[k\pi t]} & \text{ if } k \text{ even} \\
\sqrt{2} \cos{[(k-1)\pi t]} & \text{ if } k \text{ odd and } k > 1
\end{array} \right. $$ 

\noindent Three distributions for the $v_{i,j,k}, j = 1,2$ were considered: (i) $v_{i,j,k} \sim \mathcal{N}(0,1)$, (ii) $v_{i,j,k} \sim t(4)/\sqrt{2}$ and (iii) $v_{i,j,k} \sim (\chi^2(4) - 4)/(2\sqrt{2})$.

\noindent Three types of clusters were simulated with an intensity controlled by a parameter $\alpha > 0$. The three chosen shifts $\Delta$ vary over time and are positive and non-zero (except possibly when $t = 0$ or $t = 1$): $\Delta_1(t) = \alpha t$, $\Delta_2(t) = \alpha t (1-t)$ and $\Delta_3(t) = \alpha \exp{[ - 100 (t-0.5)^2 ]}/3$. Thus, these clusters correspond to both magnitude clusters and shape clusters. Different values of the true cluster intensity were considered for each $\Delta$. Since $\Delta_2$ and $\Delta_3$ takes lower values over time than $\Delta_1$ for the same value of $\alpha$, it needs greater $\alpha$ values. 
Thus, $\alpha \in \{ 0 ; \ 0.75 ; \ 1.5 ; \ 2.25 ; \ 3 \}$ for $\Delta_1$, $\alpha \in \{ 0 ; \ 2 ; \ 4 ; \ 6 ; \ 8 \}$ for $\Delta_2$ and $\alpha \in \{ 0 ; \ 2.5 ; \ 5 ; \ 7.5 ; \ 10 \}$ for $\Delta_3$.
It is noteworthy that $\alpha = 0$ was also tested in order to evaluate the maintenance of the nominal type I error.
An example of the data for the Gaussian distribution for the $v_{i,j,k}$ is given in the Appendix (Figure \ref{fig:delta2alpha8}).

\subsubsection{Comparison of the methods}

For each type of process, each type of $\Delta$, and each value of $\alpha$, we simulated 1000 artificial datasets. The p-value associated with each MLC was estimated by generating 999 random permutations of the data. A type I error of 5\% was considered for the rejection of $\mathcal{H}_0$. The three methods’ respective performances were compared with regard to four criteria: the power, the true positive rate, the false positive rate, and the F-measure. 

\noindent The power was defined as the proportion of simulations leading to the rejection of $\mathcal{H}_0$, according to the type I error. For the simulated datasets yielding to the rejection of $\mathcal{H}_0$, the true positive rate is the mean proportion of sites correctly detected among the sites in $w$, the false positive rate is the mean proportion of sites in  $w^\mathsf{c}$ that were included in the detected cluster, and the F-measure corresponds to the average harmonic mean of the proportion of sites in $w$ within the detected cluster (positive predictive value) and the true positive rate.

\subsubsection{The results of the simulation study}

\begin{figure}
\begin{center}
	\includegraphics[width=1\textwidth]{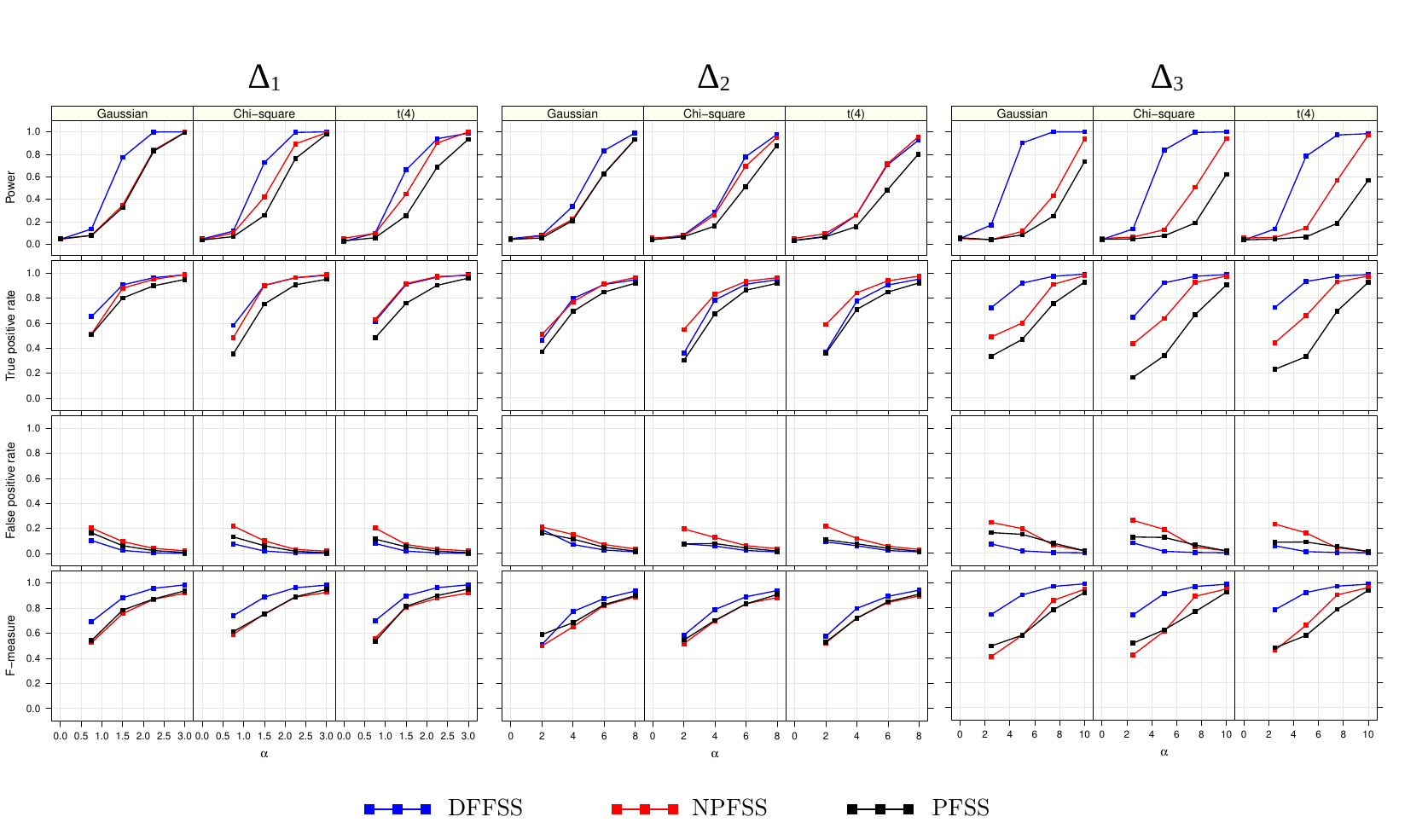}
	\caption{The simulation study: comparison of the NPFSS, DFFSS and PFSS methods for three types of shift $\Delta$: $\Delta_1(t) = \alpha t$ (left panel), $\Delta_2(t) = \alpha t (1-t)$ (middle panel) and $\Delta_3(t) = \alpha \exp{[ - 100 (t-0.5)^2 ]}/3$ (right panel). For each method, the power curves, the true-positive and false-positive rates, and the F-measure values for detection of the true cluster as the MLC are shown. $\alpha$ is the parameter that controls the cluster intensity.}
	\label{fig:simuresults}
\end{center}
\end{figure}

\noindent The results of the simulation are shown in Figure \ref{fig:simuresults}. When $\alpha = 0$ the different methods seem to maintain the correct type I error (0.05), regardless of the type of process and the $\Delta$ (see power curves in Figure \ref{fig:simuresults}). \\ 

\noindent The DFFSS almost always shows higher powers than the two other methods for all the scenarios, especially in the case of the local shift $\Delta_3$. While the NPFSS and the PFSS presents similar powers when the $v_{i,j,k}$ are gaussian for $\Delta_1$ and $\Delta_2$ but when the data were not normal or the shift is local ($\Delta_3$) the NPFSS shows better powers than the PFSS. \\
Generally, as expected since it is parametric, the PFSS performed less well when the data were not distributed normally. Even if it is presented as ``distribution-free'', the DFFSS is based on a Student test statistic which works better for normal data: we can see that its performances also slightly decrease when the $v_{i,j,k}$ follow a Student distribution.
The DFFSS and the NPFSS show higher true positive rates than the PFSS, for the shift $\Delta_2$ the NPFSS also presents slightly higher true positive rates than the DFFSS. However the true positive rates of the DFFSS are much higher than the NPFSS for the local shift $\Delta_3$. In terms of false positives, the DFFSS always presents the lowest false positive rates. The NPFSS presents the highest false positive rates, yielding to low F-measures. In contrast the DFFSS presents the highest F-measures. The PFSS also shows quite low F-measures which is due to its low true positive rates. However its F-measures are often higher than for the NPFSS. 

\section{Application to real data} \label{sec:realdata}

\subsection{Unemployment rates in France}

\begin{figure}[!h]
\begin{center}
	\includegraphics[width = 1\textwidth]{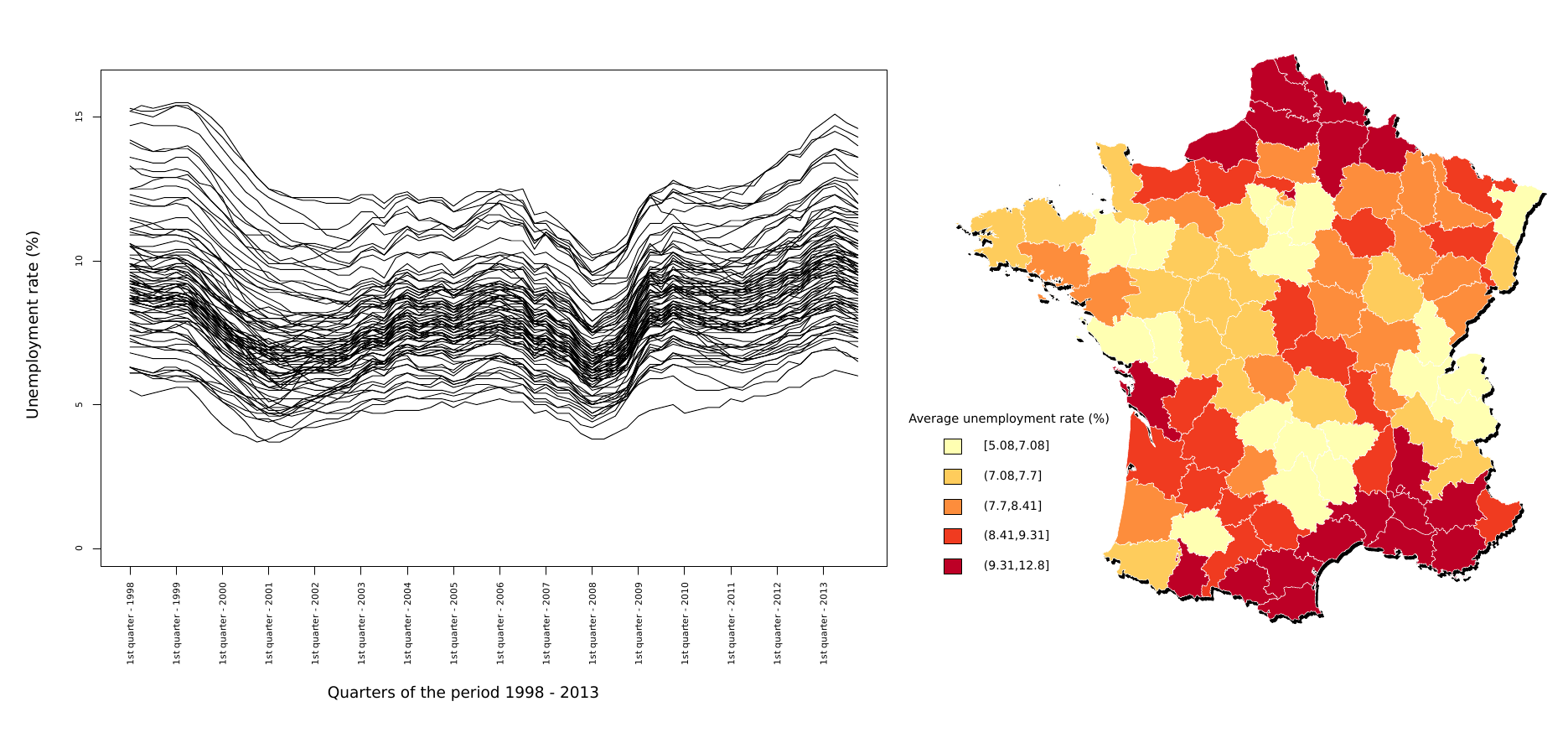}
	\caption{Unemployment rate curves from 1998 to 2013 in each of the 94 French \textit{départements} (left panel), and the spatial distribution of the mean unemployment rates over the period (right panel).}
	\label{fig:courbes_chomage}
\end{center}
\end{figure}

\noindent We considered data on unemployment rates in France; these were provided by the \textit{Institut National de la Statistique et des Etudes Economiques} (Paris, France) for each quarter between 1998 and 2013 (64 values) and for each of the 94 French \textit{départements} (Figure \ref{fig:courbes_chomage}, left panel). It can be seen that the unemployment rates varied markedly over time. Furthermore, the spatial distribution of the mean unemployment rate between 1998 and 2013 (Figure \ref{fig:courbes_chomage}, right panel) was heterogeneous. High unemployment rates tended to aggregate in northern France and south-eastern France. These two observations suggested that functional spatial scan statistics could be used to detect \textit{départements}-level spatial clusters of unemployment.

\subsection{Spatial clusters detection}

\noindent In order to detect spatial clusters of low or elevated unemployment rates, we used the PFSS, the DFFSS and the NPFSS to detect the MLC and secondary clusters that had a high value of the concentration index ($F_n^{(w)}$ for PFSS, $I^{(w)}$ for DFFSS, and $U(w)$ for NPFSS; see \cite{wilco_cucala} for details) and that did not cover the MLC \citep{spatialscanstat}. For both methods, we considered a circular variable scanning window whose maximum size was set so that a cluster never contained more than half the total number of \textit{départements}. The statistical significance of the MLC and the secondary clusters was evaluated in 999 Monte-Carlo permutations. A spatial cluster was considered to be statistically significant when the associated p-value was below 0.05.

\subsection{Results}

\begin{figure}[h!]
\begin{center}
	\includegraphics[width = 1\textwidth]{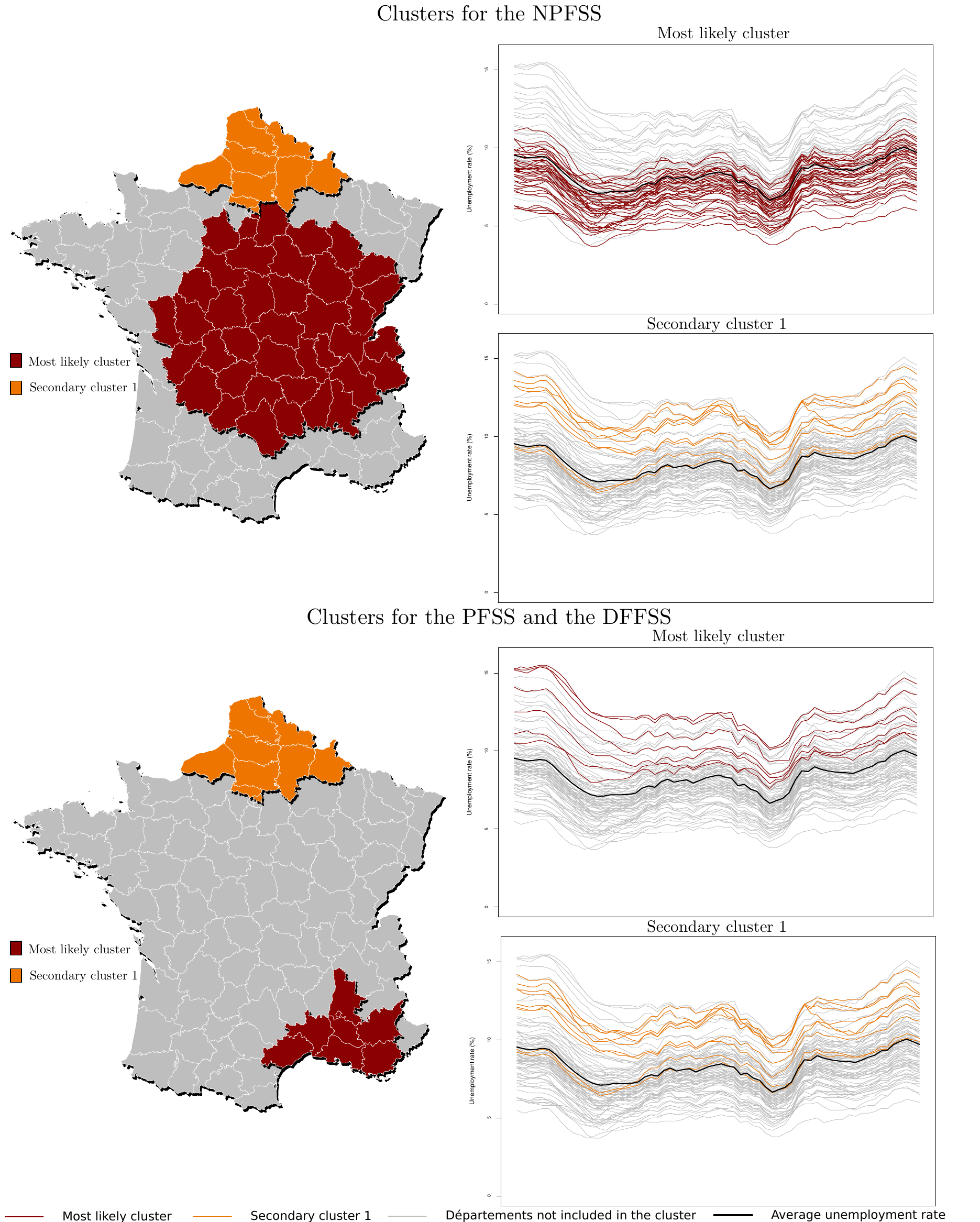} 
	\caption{Significant spatial clusters of high or low unemployment rates detected by the NPFSS (top panel) and the PFSS and the DFFSS (bottom panel). For each method, the MLC is shown in red and the secondary cluster is shown in orange. The unemployment rate curves (from 1998 to 2013) for the \textit{département} in the MLC and in the secondary cluster are presented as correspondingly coloured lines in the graphs. The black curve is the national average unemployment rate.}
	\label{fig:noconstraint}
\end{center}
\end{figure}

\noindent The DFFSS, the PFSS and the NPFSS methods each detected two significant spatial clusters (Figure \ref{fig:noconstraint} and Table \ref{table:noconstraint}).
\noindent  The PFSS and the DFFSS identified the same two significant spatial clusters of elevated unemployment rates: the MLC (7 \textit{départements}, $\hat{p}_{\text{PFSS}}=0.011$,  $\hat{p}_{\text{DFFSS}}=0.005$) was located in south-eastern France and the secondary cluster (9 \textit{départements}, $\hat{p}_{\text{PFSS}}=0.039$,  $\hat{p}_{\text{DFFSS}}=0.026$) was located in northern France. These clusters were homogeneous because they contained only curves that were above the national average unemployment rate (except for two \textit{départements} in the secondary cluster at the beginning of the period studied).\\
The NPFSS identified a significant large (40-\textit{départements}) MLC  ($\hat{p}_{\text{NPFSS}}=0.013$) with low unemployment rates in the centre of France. The unemployment rate curves were heterogeneous, and the MLC included some \textit{départements} with an unemployment rate above the national average (Figure \ref{fig:noconstraint}). The secondary cluster (9 \textit{départements}, $\hat{p}_{\text{NPFSS}}=0.039$) detected by the NPFSS was exactly the same as that detected by the PFSS and the DFFSS.

\begin{table}[!h]
\centering
\caption{Statistically significant spatial clusters of higher or lower unemployment rates detected using the NPFSS, the PFSS and the DFFSS}
\begin{tabular}{lccc}
	\hline
	& \textbf{Cluster} & \textbf{\# \textit{départements}} & \textbf{p-value} 
	
	\\ \hline
	\textbf{NPFSS}  & 1  & 40 & 0.013 
	\\
	& 2 & 9  &  0.039  
	\\ \hline
	\textbf{PFSS} & 1 & 7 & 0.011 
	\\
	& 2  & 9   & 0.039 
	\\ \hline
	\textbf{DFFSS} & 1 & 7 & 0.005 
	\\
	& 2  & 9   & 0.026 
	\\ \hline
\end{tabular}
\label{table:noconstraint}
\end{table}

\section{Discussion} \label{sec:discussion}
\noindent Here, we developed a PFSS and a DFFSS for detecting clusters of functional data indexed in space. They are respectively based on the adaptation of the ANOVA F-statistic initially developed by \cite{anova_functional} and the ``distribution-free'' scan statistic of \cite{a_distribution_free} as a concentration index for spatial scan statistics. For functional spatial data, our methods appear to be more relevant than univariate spatial scan statistics because the latter data average data over a time period and thus lose a large amount of information. Furthermore, our methods appear to be more relevant than multivariate spatial scan statistics because it deals with two key problems: high dimensionality, and strong correlation between time measurements.\\

\noindent In a simulation study, we compared our new methods with the NPFSS developed by \cite{wilco_cucala}.
The results of the simulation study showed that all the methods maintain the nominal type I error. In case of normally distributed data, the PFSS performed better than the NPFSS: the two methods had similar powers (except for $\Delta_3$) but the NPFSS tended to detect larger clusters. The PFSS has the advantage of detecting smaller clusters, which might be more relevant in many applications. Moreover, the F-measure value was almost always higher for the PFSS than for the NPFSS, which thus increase the level of confidence in the detected clusters.
For all distributions the DFFSS appears to perform better than the NPFSS and the PFSS. Moreover it should be noted than it also presents a better power to detect spatial clusters that appear in a short period of time ($\Delta_3$ case in the present simulation study). 
\\

\noindent We came to the same conclusions by studying the application to real data. The PFSS, the DFFSS and the NPFSS were used to detect clusters of unemployment rates in the 94 French \textit{départements}. The NPFSS detected an MLC composed of 40 \textit{départements} in the centre of France. Although this was a ``low unemployment cluster'', it contained many \textit{départements} with high unemployment rate curves and so was not relevant – especially since the cluster covered approximately half of France. In comparison, the MLC for the PFSS and the DFFSS was the same and was only composed of 7 \textit{départements} in south-eastern France, and all of these had high unemployment rates. All approaches detected the same secondary cluster of high unemployment rates in northern France.\\

\noindent It should be noted that the DFFSS and the PFSS applied here dealt solely with round-shaped clusters. In the application to unemployment data, an elongated cluster of \textit{départements} of high unemployment rates was found in southern France – suggesting that other cluster shapes should be considered. In fact, the DFFSS and the PFSS can easily be adapted to other cluster shapes, such as elliptic clusters \citep{elliptic} and graph-based clusters \citep{cucala_graph}.\\

\noindent It is also noteworthy that in the simulation study and the study of real data, the observation times were identical at each spatial location. Although this is an ideal situation for functional data, it does not apply to many real datasets. However, our new methods can be still applied to the case in which observation times are not identical at each spatial location – notably by using well-known methods (e.g. projection into a B-spline basis) to smooth the raw data (for details, see \cite{ramsay2005principal}).\\

\noindent The PFSS is derived from the ANOVA test for functional data introduced by \cite{anova_functional}.  More recently, \cite{srivastava} developed another two-sample test statistic for functional data. However, the latter appears to be very time consuming and we did not consider it for the present spatial scan statistic. In fact, the concentration index is calculated first for each potential MLC and then again when computing the latter’s statistical significance (i.e. the detection step is performed for each permutation of the data). Thus, we considered Cuevas’ approach because it yields a reasonable computing time for the spatial scan statistic.\\

\noindent Although the DFFSS and the PFSS were designed to handle univariate functional data, the multivariate functional case should also be investigated. By way of an example, one could consider the data from sensors located in different geographical locations (e.g. sensors simultaneously measuring several air pollutants) over time. In such a case, the detection of significant spatial clusters might help experts to identify environmental black spots. The PFSS could be extended to the multivariate functional framework by adapting the multivariate ANOVA for functional data suggested by \cite{manovafonctional}. 
It should be noted that the functional Wilcoxon-Mann-Whitney test developed by \cite{wilco} can also be adapted for use with multivariate functional data by taking account of a suitable scalar product for calculation of the sign function. \\

\newpage
\bibliographystyle{model5-names}
\bibliography{bibliographie}

\appendix

\section{Supplementary materials}\label{simulated_cluster}
\begin{figure}[!h]
\centering
\includegraphics[width = 0.5\textwidth]{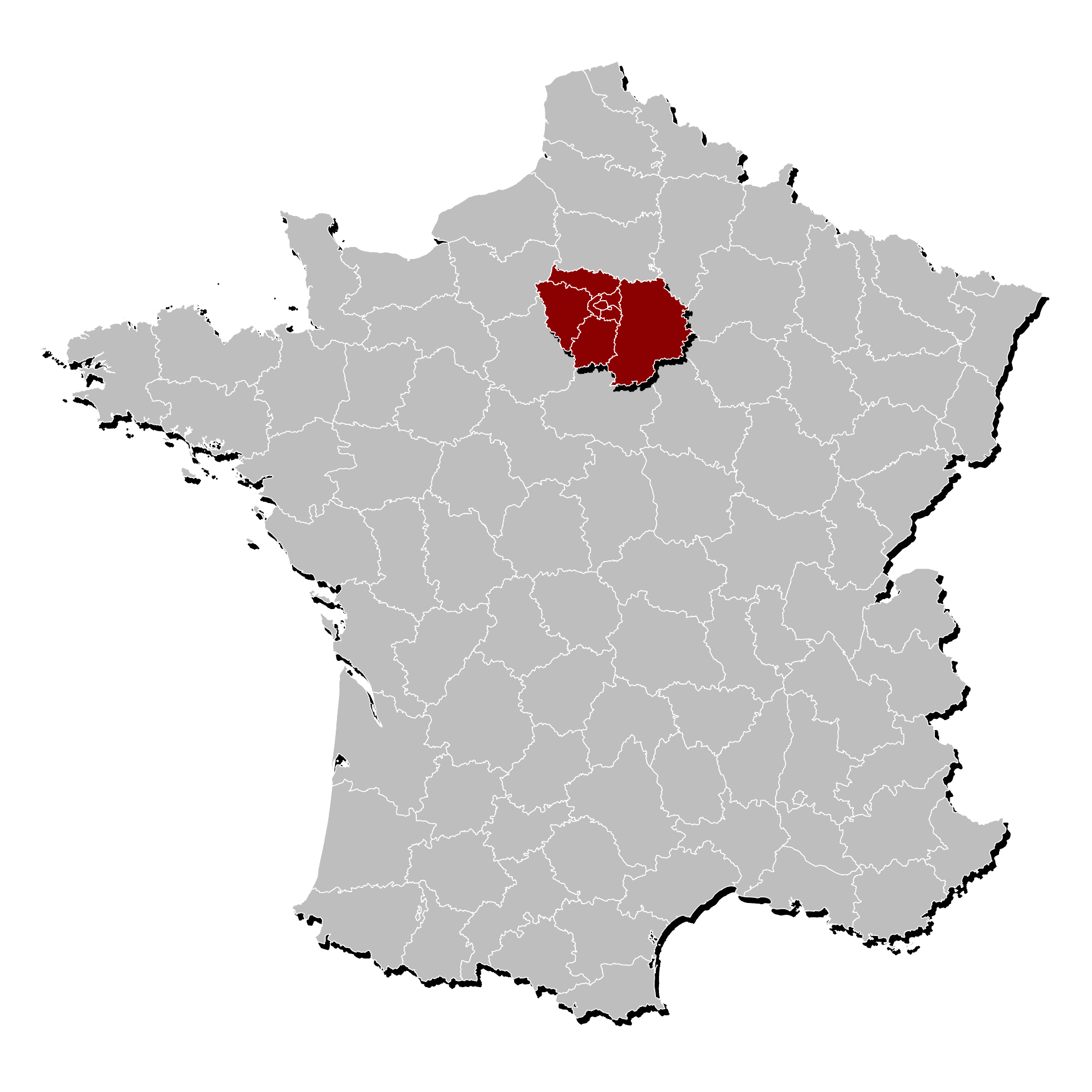}
\caption{The 94 French \textit{départements} and the true cluster (in red) simulated for each artificial dataset.}
\label{fig:sitesgrid}
\end{figure}

\begin{figure}[!h]
\begin{center}
	\includegraphics[width = \linewidth]{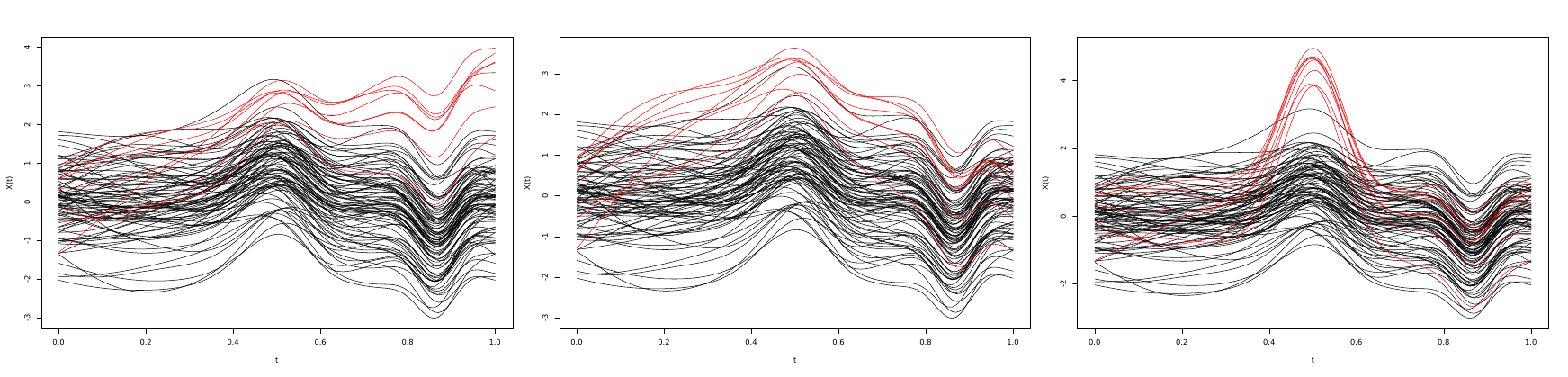}
	\caption{The simulation study: an example of the data generated for the Gaussian process, with $\Delta(t) = \Delta_1(t) = 3t$ (left panel), $\Delta(t) = \Delta_2(t) = 8t(1-t)$ (middle panel) and $\Delta(t) = \Delta_3(t) = 10 \exp{[ - 100 (t-0.5)^2 ]}/3$ (right panel). The red curves correspond to the observations in the cluster.}
	\label{fig:delta2alpha8}
\end{center}
\end{figure}

\noindent Figure \ref{fig:delta2alpha8} shows an example of the data generated when the $v_{i,j,k}$ are Gaussian.

\section{Reduction of the computation time for the NPFSS}
\label{Reduction_computing_time}
\noindent \cite{wilco_cucala} defined a nonparametric scan statistic for independent observations of a process in a separable Banach space $\chi$, $\mathcal{S}_{\mathcal{W}}$ by 
$$\mathcal{S}_{\mathcal{W}} = \underset{w \in \mathcal{W}}{\max} \ \bigg| \bigg| \dfrac{1}{\sqrt{|w| \ |w^\mathsf{c}| \ (|w| + |w^\mathsf{c}|)}} \sum_{i, s_i \in w} \sum_{j, s_j \in w^\mathsf{c}} \text{sign}(X_j - X_i) \bigg| \bigg| $$ \\
where the sign function is the Gâteaux-derivative of the norm on $\chi$.

\subsection*{Calculation of the sign function in the case of square-integrable functions}

\noindent Let $X \in L^2(\mathcal{T}, \mathbb{R}) \textbackslash \{0\}$. In this section, we shall compute the $\text{sign}(X)$ which is the Gâteaux-derivative of $||X||_2 = \sqrt{\langle X,X \rangle}$ where $\langle X,Y \rangle = \int_{\mathcal{T}} X(t) Y(t) \ \text{d}t$. \\

\noindent Let $h \in L^2(\mathcal{T}, \mathbb{R})$,\\
\begin{flalign*}
\underset{v \rightarrow 0}{\lim} \ \dfrac{||X + hv||_2 - ||X||_2}{v} &= \underset{v \rightarrow 0}{\lim} \ \dfrac{||X + hv||_2^2 - ||X||_2^2}{v(||X + hv||_2 + ||X||_2)} &\\[0.2cm]
&= \underset{v \rightarrow 0}{\lim} \ \dfrac{\langle X+hv, X+hv \rangle - \langle X, X \rangle}{v(||X + hv||_2 + ||X||_2)} &\\[0.2cm]
&= \underset{v \rightarrow 0}{\lim} \ \dfrac{2v \langle h, X \rangle + v^2 ||h||_2^2}{v(||X + hv||_2 + ||X||_2)} &\\[0.2cm]
&= \underset{v \rightarrow 0}{\lim} \ \dfrac{2 \langle h, X \rangle + v ||h||_2^2}{||X + hv||_2 + ||X||_2} &\\[0.2cm]
&= \dfrac{\langle h, X \rangle}{||X||_2}.
\end{flalign*}

\noindent Then $\text{sign}(X)(h) = \dfrac{\langle h, X \rangle}{||X||_2}$ if $X \in L^2(\mathcal{T}, \mathbb{R}) \textbackslash \{0\}$.

\subsection*{Reduction of the computation time}

\noindent Let $X$ be a stochastic process taking values in $L^2(\mathcal{T}, \mathbb{R})$, and $X_1, \dots, X_n$ be the realisations of $X$ in the locations $s_1, \dots, s_n$. \\
Since the $X_i$ are in $L^2(\mathcal{T}, \mathbb{R})$,  $\text{sign}(X_i) = \left\{ 
\begin{array}{cl}
\dfrac{X_i}{||X_i||_2} & \text{ if } X_i \neq 0 \\
0   &  \text{ if } X_i = 0
\end{array}
\right.$. \\

\noindent Then:
\begin{flalign*}
\sum_{\substack{i=1 \\ s_i \in w}}^n \sum_{\substack{j=1 \\ s_j \in w^\mathsf{c}}}^n \text{sign}(X_j - X_i) &= \sum_{\substack{i=1 \\ s_i \in w}}^n \sum_{j=1}^n \text{sign}(X_j - X_i) - \sum_{\substack{i=1 \\ s_i \in w}}^n \sum_{\substack{j=1 \\ s_j \in w}}^n \text{sign}(X_j - X_i) &\\
&= \sum_{\substack{i=1 \\ s_i \in w}}^n \sum_{j=1}^n \text{sign}(X_j - X_i) - \sum_{\substack{i=2 \\ s_i \in w}}^n \sum_{\substack{j=1 \\ s_j \in w}}^{i-1} \text{sign}(X_j - X_i) - \sum_{\substack{i=1 \\ s_i \in w}}^{n-1} \sum_{\substack{j=i+1 \\ s_j \in w}}^n \text{sign}(X_j - X_i) &\\
&= \sum_{\substack{i=1 \\ s_i \in w}}^n \sum_{j=1}^n \text{sign}(X_j - X_i) - \sum_{\substack{i=2 \\ s_i \in w}}^n \sum_{\substack{j=1 \\ s_j \in w}}^{i-1} \text{sign}(X_j - X_i) + \sum_{\substack{j=1 \\ s_j \in w}}^{n-1} \sum_{\substack{i=j+1 \\ s_i \in w}}^n \text{sign}(X_j - X_i) &\\
&= \sum_{\substack{i=1 \\ s_i \in w}}^n \sum_{j=1}^n \text{sign}(X_j - X_i) - \sum_{\substack{i=2 \\ s_i \in w}}^n \sum_{\substack{j=1 \\ s_j \in w}}^{i-1} \text{sign}(X_j - X_i) + \sum_{\substack{i=2 \\ s_i \in w}}^n \sum_{\substack{j=1 \\ s_j \in w}}^{i-1} \text{sign}(X_j - X_i) &\\
&= \sum_{\substack{i=1 \\ s_i \in w}}^n \sum_{j=1}^n \text{sign}(X_j - X_i)
\end{flalign*}

\noindent Thus the matrix of signs (in which each line $i$ corresponds to $\sum_{j=1}^n \text{sign}(X_j - X_i)$) can be computed. Lastly, to obtain $\sum_{\substack{i=1 \\ s_i \in w}}^n \sum_{\substack{j=1 \\ s_j \in w^\mathsf{c}}}^n \text{sign}(X_j - X_i)$, one sums the rows of the matrix that correspond to the sites in $w$. \\


\noindent In order to evaluate the order of magnitude of the reduction in computing time, let consider the following example: 1000 datasets of the previous simulation were simulated. 
The improved method also took into account the computation time of the matrix of signs. The computation times for the two methods are shown in Table \ref{tab:boxplot_temps}. In this example, our method divided the computing time by a factor of approximately 100.  

\begin{table}[!h]
\caption{Comparison of the computation times (in seconds) in a simulation before and after the improvement with the matrix of signs. The results are for 1000 simulated datasets 
	observed at 101 equally spaced times of $[0;1]$. 
}
\centering
\begin{tabular}{l c c}
	\hline
	\textbf{Method} & \textbf{Range} & \textbf{Mean (sd)} \\
	\hline
	\textbf{Basic method} &  [78 ; 120] & 101 (8)\\
	\textbf{Improved method} & [0.25 ; 0.87] & 0.34 (0.06) \\
	\hline
\end{tabular}
\label{tab:boxplot_temps}
\end{table}

\end{document}